%% file: main.tex
\documentclass[%
 reprint,
superscriptaddress,
 amsmath,amssymb,
 aps,
 lengthcheck,
]{revtex4-2}

\usepackage{graphicx}
\usepackage{soul}
\usepackage{dcolumn}
\usepackage{bm}
\usepackage[usenames,dvipsnames]{xcolor}
\usepackage[colorlinks=true, urlcolor=blue]{hyperref}
\DeclareMathAlphabet\mathbfcal{OMS}{cmsy}{b}{n}
\usepackage[english]{babel}
\usepackage{bm}

\begin{document}

\preprint{APS/123-QED}

\title{Mapping the limits of equilibrium in sheared granular liquid crystals}
\thanks{Published as \textit{Physical Review Research, 8, 043151 (2026)}\\
DOI: \url{https://doi.org/10.1103/cjdt-6d9p}}

\author{Jacopo Bilotto}
\email{jacopo.bilotto@epfl.ch}
\affiliation{Institute of Civil Engineering, Institute of Materials Science and Engineering, École Polytechnique Fédérale de Lausanne (EPFL), CH 1015 Lausanne, Switzerland}

\author{Martin Trulsson}
\affiliation{Computational Chemistry, Lund University, Lund SE-221 00, Sweden}

\author{Jean-François Molinari}
\affiliation{Institute of Civil Engineering, Institute of Materials Science and Engineering, École Polytechnique Fédérale de Lausanne (EPFL), CH 1015 Lausanne, Switzerland}

\date{November 10, 2026}

\begin{abstract}
While elongated particles in viscous fluids follow Jeffery orbits, their dynamics in dense granular flows remain unresolved.
We demonstrate that sheared, low friction granular rods do not follow these orbits, but instead reach a quasi-equilibrium state quantitatively described by classical liquid crystal theory, with collisions providing effective thermal noise.
This quasi-equilibrium analogy breaks down at two limits: low aspect ratios, where the theory incorrectly predicts isotropy, and high inter-particle friction, which drives a transition from steric screening to frictional gearing. 
We map this far-from-equilibrium breakdown using a phenomenological effective Ericksen number, establishing the limits of thermal thermodynamic analogies in driven athermal matter.
\end{abstract}

\maketitle


\section{Introduction}

Granular materials, though athermal and dissipative, can surprisingly arrange into ordered, equilibrium-like configurations when driven \cite{reis_crystallization_2006}.
Elongated grains, for instance, exhibit isotropic–nematic transitions driven purely by geometric steric exclusion \cite{galanis_spontaneous_2006, galanis_nematic_2010, borzsonyi_granular_2013} and align in shear flows \cite{borzsonyi_orientational_2012, wegner_alignment_2012, berzi_stresses_2016, berzi_collisional_2017, trulsson_rheology_2018, nagy_flow_2020, bilotto_shear_2025a}, even at modest shape anisotropies \cite{marschall_orientational_2019}.  
This behavior strongly resembles the isotropic-to-nematic transition of thermal liquid crystals, which is well described by equilibrium Onsager statistical mechanics in dilute systems \cite{onsager_effects_1949} and by Parsons's \cite{parsons_nematic_1979} and Lee's \cite{lee_numerical_1987} correction for high density.
The non-equilibrium driving is also effectively captured in the Doi--Edwards--Kuzuu (DEK) framework \cite{doi_dynamics_1978, doi_molecular_1981, kuzuu_constitutive_1984}, which quantitatively describes these thermal systems under flow.
Although attempts have been made to model the ordering of granular monolayers \cite{muller_ordering_2015}, a corresponding statistical tool is currently lacking for athermal systems.
Treating collisions as an effective source of thermal noise \cite{talbot_exploring_2024} has emerged as a promising pathway to formulate an out-of-equilibrium theory.
Using classical equilibrium theory, we show that this approach accurately predicts the orientation distribution of elongated grains, provided the grains are sufficiently elongated and with low friction.
However, the analogy breaks down when the non-equilibrium rotational driving overwhelms the steric caging.
In our specific system this divergence manifests both at low anisotropies where steric caging is inherently weak, and at high interparticle friction.
At higher elongations, crossing this threshold triggers a transition from a quasi-equilibrium steric ``screening'' state to a far-from-equilibrium frictional ``gearing'' state.

\section{Methods}

To investigate the alignment of athermal granular systems under shear, we employ a Discrete Element Method (DEM) with soft-core Hertz-Mindlin frictional interactions, a Coulomb friction coefficient $\mu_p$, and a normal restitution coefficient of 0.1.
All systems consist of $N=2000$ spheroids with a $20\%$ degree of polydispersity\footnote[3]{Simulations with no polydispersity showed little difference in the results}, and are sheared through Lees-Edwards boundary conditions \cite{lees_computer_1972} at constant normal stress $\sigma_{yy}$, imposed along the gradient direction, maintaining an inertial number $I = \dot{\gamma} d \sqrt{\rho/\sigma_{yy}} \approx 0.1$, where $\dot{\gamma}$ is the shear rate, $d$ the particle equivalent sphere diameter, and $\rho$ its density.
We have checked that an almost linear velocity profile is achieved for all simulations.
The contact stiffness number $\kappa = (E/\sigma_{yy})^{2/3} > 2  \cdot 10^3$, with $E$ the grains' Young's modulus, is set sufficiently high to ensure the overlap between grains is small.
A snapshot of the simulation set-up is shown in Fig.~\ref{fig:setup_S2_phi}a.
Orientation statistics are calculated by averaging over all particles in steady state, see SM~\ref{app:steady_state}~\cite{SupplementalMaterial}, extracted after an accumulated strain of $\gamma = 20$, and time-averaged over an additional $10$ strain units ($1000$ snapshots) yielding $2 \times 10^6$ distinct particle configurations per run.
Further simulation details are provided in \cite{bilotto_shear_2025a}, and robustness checks in SM~\ref{app:rob_check}~\cite{SupplementalMaterial}.
While our primary parametric study employs spheroids (as depicted in Fig.~\ref{fig:setup_S2_phi}), we performed an additional, targeted set of simulations using rod-like superellipsoids (detailed in SM~\ref{app:theory_experiments}~\cite{SupplementalMaterial}) to accurately validate our model against the cylindrical grains used in experimental data \cite{wegner_alignment_2012, borzsonyi_shearinduced_2012, borzsonyi_orientational_2012, wegner_effects_2014}.
These validation runs followed the same protocol as the spheroid simulations.
Experimental comparisons are made with data from these references, which were obtained with X-ray CT scans in split-bottom shear cell measurements and averaged over long strains in the shearing zone.

\section{Mathematical framework}

With this physical system established, we look for the appropriate mathematical framework to describe the steady-state orientational statistics.
We test whether the continuum framework originally developed for thermal nematic suspensions can be adapted for driven athermal grains.\\
Consider an axisymmetric particle with aspect ratio $\alpha>1$ in a simple shear flow at high packing fraction. 
To model how these particles collectively align, we track their orientation distribution $f(\mathbf{u})$, where $\mathbf{u}$ is the unit vector defining the orientation of the particle's major axis.
The preferred macroscopic orientation is the director $\mathbf{n}$ (principal eigenvector of $\langle \mathbf{u} \otimes \mathbf{u} \rangle$), lying in the flow-gradient plane at an angle $\eta$ to the flow.
Particle orientations are described by the polar angle from the director, $\theta = \arccos(\mathbf{u} \cdot \mathbf{n})$, and the in-plane alignment angle, $\vartheta = \arctan(u_y/u_x)$, see inset of Fig.~\ref{fig:setup_S2_phi}a.
Following DEK theory \cite{doi_molecular_1981, larson_arrested_1990}, the evolution of this distribution is governed by a Smoluchowski-type equation, which balances the rotation driven by the macroscopic flow against the effective diffusion caused by inter-particle collisions. 
In athermal systems under simple shear flow, the rotational diffusion $D_r$ scales linearly with shear rate $\dot \gamma$ (see SM~\ref{app:rot_diffusion}~\cite{SupplementalMaterial}), such that $D_r =  \tilde{D}_r(f(\mathbf{u})) \dot \gamma$, where the mobility $\tilde{D}_r(f)$ is, in our case, derived from extended tube models \cite{chen_scaling_2020}. 
In these models, the dense surrounding neighbors act as a tight physical cage, the ``tube'', severely restricting the particle's ability to rotate.
Additionally, solid contacts \cite{pol_collective_2026} alter the deterministic torque relative to the dilute Jeffery orbit \cite{jeffery_motion_1922}.
While closed-form microscopic expression for the drift remains an open challenge \cite{quinones_smoluchowski_2026}, our general framework might in principle apply to different classes of interactions, of different physical origins.
We define this effective driving flux as $\mathbfcal{J}(\mathbf{u})$. 
Eliminating time in favor of accumulated strain $\gamma$ \cite{borzsonyi_shearinduced_2012}, the evolution equation is:
\begin{align}
    {\partial f \over \partial \gamma} + 
    \nabla_\mathbf{u} \cdot \left[ \tilde{\mathbfcal{J}} f \right]
     = \tilde{D}_r(f)\nabla_\mathbf{u} \cdot 
    \left[ \nabla_\mathbf{u} f + f \nabla_\mathbf{u} \mathcal{U}_{\text{ex}} \right] \, ,
    \label{eq:full_smoluchowski3d}
\end{align}
where $\nabla_\mathbf{u} = (\mathbf{I} - \mathbf{u} \otimes \mathbf{u}) \cdot \frac{\partial}{\partial \mathbf{u}}$ is the gradient operator on the unit sphere, $\mathbf{I}$ being the identity tensor, $\tilde{\mathbfcal{J}} = \mathbfcal{J} / \dot \gamma$ is the normalized driving and $\mathcal{U}_{\text{ex}}$ is the mean-field steric potential \cite{onsager_effects_1949, parsons_nematic_1979}. 
If the driving flux balances the restoring diffusion (in the weak shearing limit \cite{marrucci_description_1989}), the steady state ($\partial_\gamma f = 0$) recovers the extended Maier–Saupe equilibrium distribution \cite{maier_einfache_1960}:
\begin{align}
    f_{\text{eq}}(\theta) = \frac{1}{Z} \exp{(U_2 S_2 P_2(\cos \theta) + U_4 S_4 P_4(\cos \theta) ) }
    \label{eq:eq_3D}
\end{align}
where $Z$ is a normalization constant, $\theta$ is the angle with the director of the distribution, $S_{2,4} = \langle P_{2,4}(\cos \theta) \rangle$ are the second and fourth nematic order parameters respectively, and $P_{2,4}$ are the second  and fourth order Legendre polynomials.
$U_{2,4}$ are the shape-dependent steric strengths.
Here the operator $\langle ... \rangle = \int ... f(\mathbf{u}) d \mathbf{u}$ denotes the ensemble average.
To generate our theoretical predictions, these strengths are computed explicitly from the particles' excluded volume using the Parsons-Lee approach and the Berne-Pechukas \cite{berne_gaussian_1972} approximation (see SM~\ref{app:excluded_volume}~\cite{SupplementalMaterial} for the full derivation \cite{piastra_explicit_2015, vega_virial_1997} and SM~\ref{app:softness_effects}~\cite{SupplementalMaterial} for finite grain stiffness effects \cite{barker_perturbation_1967}).
Physically, this distribution describes a steady state where steric exclusion forces the particles to preferentially align along a common axis, with $S_2$ and $S_4$ quantifying the sharpness of this alignment. 
Crucially, while an equivalent thermal system would crystallize at the high volume fractions recorded here, the imposed shear continuously disrupts translational order. 
This forced shear melting sustains a flowing \cite{zeng_unified_2026}, supercooled nematic state. 
Consequently, the macroscopic driving not only forces nematic alignment at low aspect ratios (where equilibrium predicts an isotropic phase), but it also allows highly elongated grains to achieve stronger nematic ordering at high densities without freezing into a smectic or crystalline solid.

\section{Results}

\begin{figure}
    \centering
    \includegraphics[width=\linewidth]{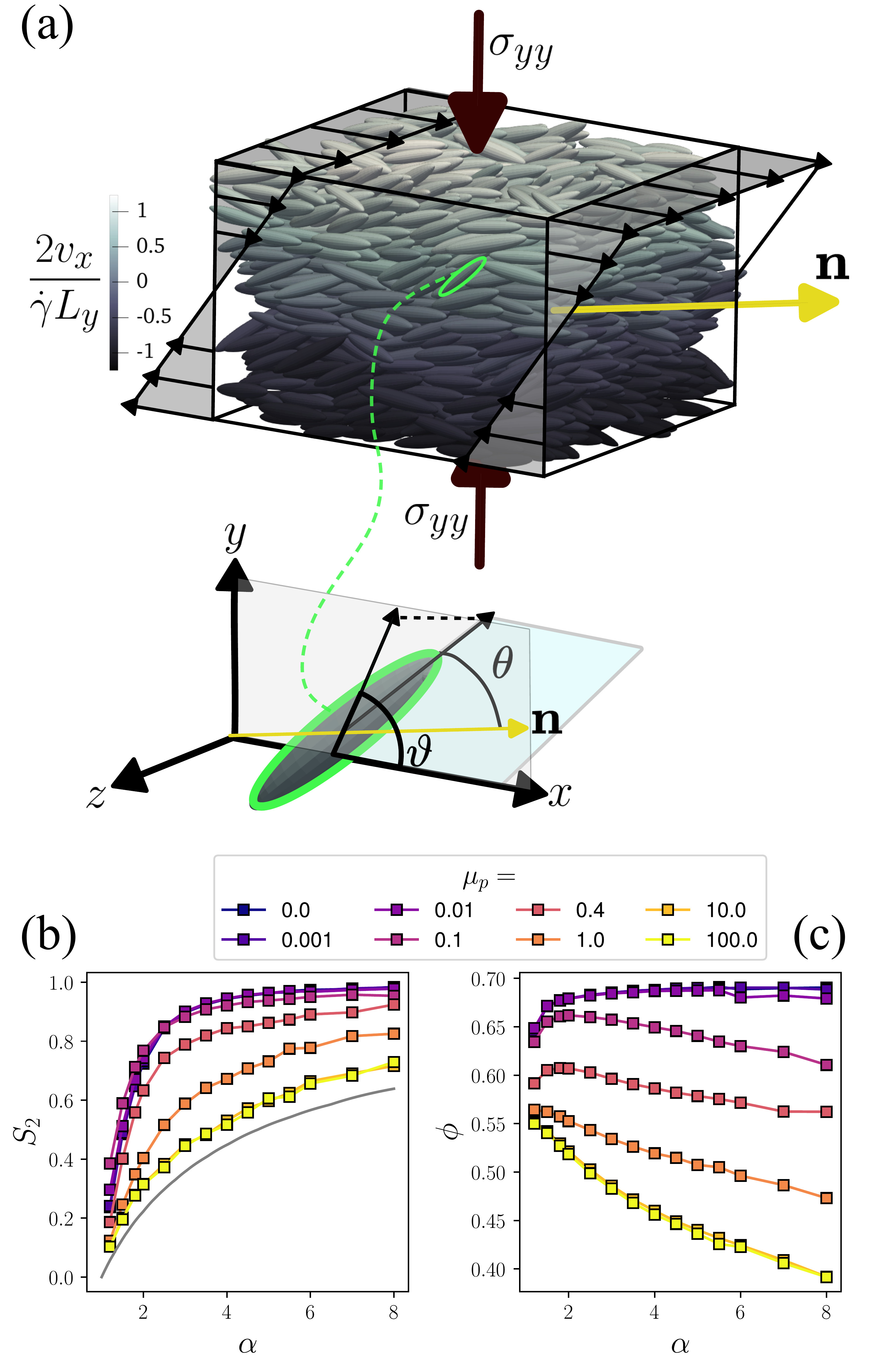}
    \caption{(a) Shear cell for $\alpha=5.0$, $\mu_p=0.4$, with particles colored by their normalized flow velocity, $v_f$, with respect to the uniform gradient $\dot \gamma L_y /2$, $L_y$ being the average box length in the gradient direction.
    A constant normal stress, $\sigma_{yy}$, is applied along the gradient direction. 
    The yellow arrow indicates the macroscopic director $\mathbf{n}$.
    For a representative particle (highlighted in green), the in-plane angle $\vartheta$ with respect to the flow and the polar angle $\theta$ with respect to the director are explicitly shown.
    (b) Nematic order parameter $S_2$ and (c) packing fraction $\phi$ as a function of aspect ratio $\alpha$.
    The plots show that increasing inter-particle friction $\mu_p$ (from dark blue to yellow) reduces nematic ordering (lowering $S_2$) and induces dilatancy (lowering $\phi$).
    The gray line marks the theoretical prediction for Jeffery orbits, without noise or steric interaction, starting from an isotropic distribution, as done in \cite{talbot_exploring_2024}. 
    All the data lie above this prediction, confirming that Jeffery orbits are not sufficient to describe the dense granular flows.
    The color scheme for $\mu_p$ is used in subsequent figures.
    \vspace{-.5cm} 
    }
    \label{fig:setup_S2_phi}
\end{figure}

With this theoretical baseline established, we can evaluate how well it captures the behavior of our highly dissipative, driven system.
Fig.~\ref{fig:setup_S2_phi}b shows the measured nematic ordering as a function of elongation and friction: as previously found \cite{trulsson_rheology_2018, nagy_flow_2020, bilotto_shear_2025a}, $S_2$ rises with $\alpha$, but friction generally reduces ordering.
This disordering correlates with a drop in average volume fraction, $\phi$, due to frictional dilatancy (Fig.~\ref{fig:setup_S2_phi}c).
Crucially, the data lie above the prediction for isolated Jeffery orbits (gray line) \cite{talbot_exploring_2024}, indicating that steric interactions are essential to maintain order in dense flows.
Additionally, the data for $\mu=10$ and $\mu_p=100$ overlap.
While these extreme values exceed standard tribological measurements, they serve to establish the theoretical infinite-friction limit, where virtually all contacts are sticking.

\begin{figure}
    \centering
    \hfill 
    \includegraphics[width=\linewidth]{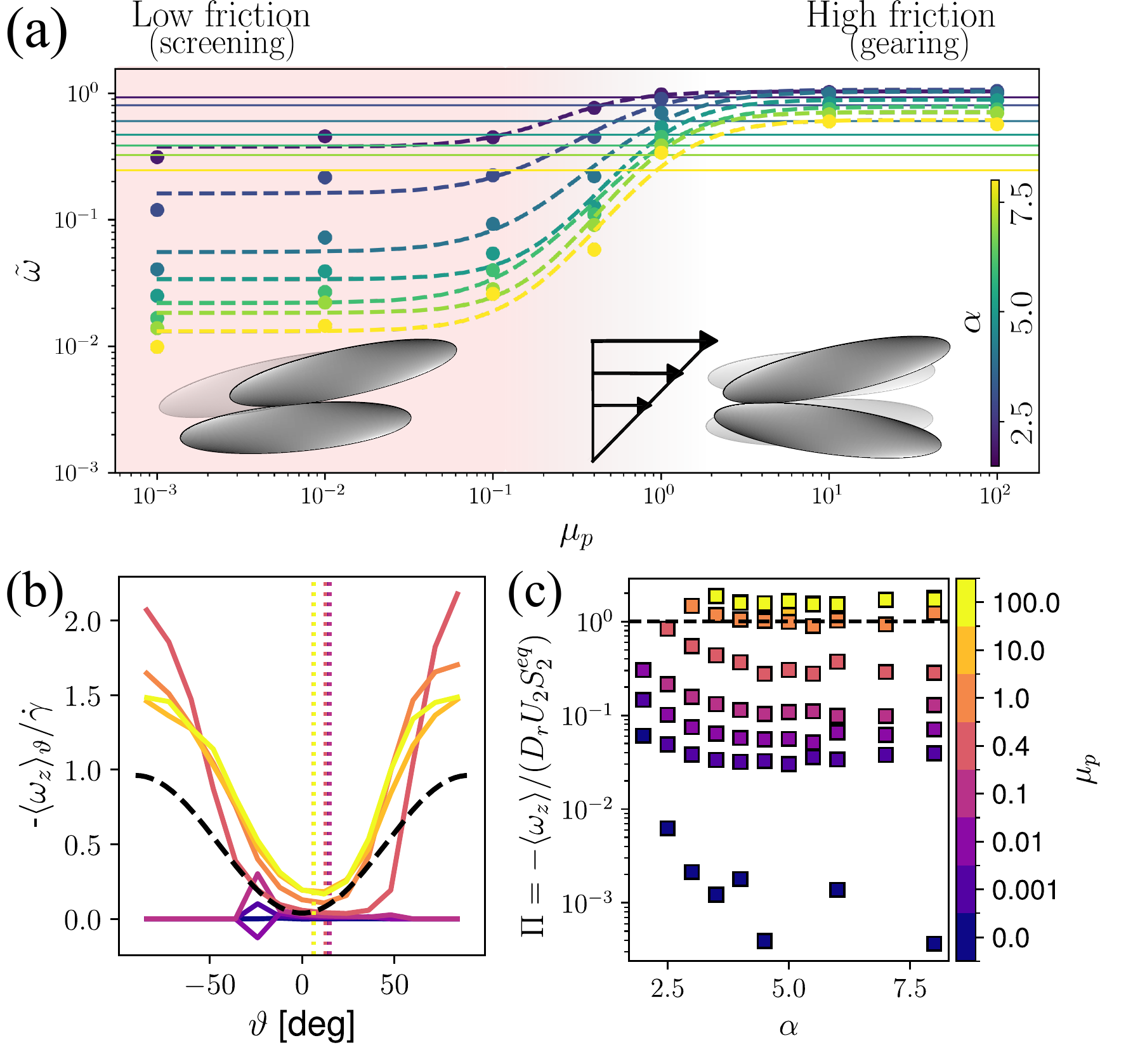}
    \caption{
    Quantifying the ``screening-to-gearing" transition.
     (a) Normalized average angular velocity $\tilde{\omega}$ extracted from simulations across varying friction coefficients (markers).
     Colors denote the particle aspect ratio $\alpha$.
     Solid lines represent the theoretical Jeffery prediction for an isolated particle in a viscous fluid, which fails to capture the dense granular dynamics. 
     Instead, the data are well-described by a two-state phenomenological model (dashed lines) capturing the transition between a low-friction ``screening'' state and a high-friction ``gearing'' state.
     In the gearing state, grains rotate faster than the isolated particle prediction, whereas in the screening state, their rotation is severely hindered.
    (b) Normalized conditional angular velocity $-\langle \omega_z\rangle_\vartheta / \dot{\gamma}$ vs. the angle with respect to the flow direction, $\vartheta$, for $\alpha=5.0$.
    The dashed black line represents the theoretical prediction for Jeffery orbits, while the vertical dotted lines indicate the measured director angle ($\eta$) in the flow-gradient plane.
    In the low-friction ``screening'' regime (dark curves, $\mu_p \le 0.01$), $\tilde \omega \approx 0$ as steric forces (director at $\eta$, dotted lines) cancel the shear torque (Jeffery orbit, black dashed line), meaning the aligned grains effectively lock into place and slide past one another without spinning.
    In the high-friction ``gearing" regime (yellow curves, $\mu_p \ge 1.0$), $\tilde \omega$ is large, showing strong rotational driving.
    (c) The effective Ericksen number $\Pi$ shows the consequence. 
    The system is in the quasi-equilibrium limit ($\Pi \ll 1$) for low friction.
    As $\mu_p$ increases, $\Pi$ grows by orders of magnitude, crossing the $\Pi=1$ threshold (dashed black line) and signaling a transition to a far-from-equilibrium, drive-dominated state.
    \vspace{-.5cm}
    }
    \label{fig:Jeffery_comparison_pi}
\end{figure}

For both simulations and experiments \cite{borzsonyi_orientational_2012, borzsonyi_shearinduced_2012}, see SM~\ref{app:steady_state}~\cite{SupplementalMaterial}, the particles reach a steady average angle of alignment towards the extension direction, with no tumbling or wagging of the director. 
Nevertheless, a stable director simply indicates a time-invariant orientational distribution; individual particles may still undergo continuous rotation \cite{talbot_exploring_2024}, yielding a non-zero mean angular velocity.
Equation \ref{eq:full_smoluchowski3d} with $\tilde{\mathbfcal{J}}$ corresponding to the Jeffery operator would erroneously predict tumbling at low shear rates, or alignment towards the compression direction at high shear rates \cite{larson_arrested_1990}. 
This motivates us to find an empirical expression for the driving, by directly measuring it from our simulations, quantified by the mean angular velocity $\langle \omega_z \rangle$ in the vorticity direction.
We observe that the suppression of order coincides with a transition in rotational dynamics.
By measuring the average normalized angular velocity $\tilde{\omega} = -2\langle \omega_z \rangle / \dot \gamma$, we identify two distinct regimes (Fig.~\ref{fig:Jeffery_comparison_pi}a).
In the frictionless limit ($\mu_p \to 0$), $\tilde{\omega}$ vanishes at high elongation. 
This is the ``steric screening'' state: collective nematic alignment generates a steric potential that effectively cancels the shear torque, preventing rotation (Fig.~\ref{fig:Jeffery_comparison_pi}b, dark curves showing the conditionally averaged angular velocity $\langle \omega_z \rangle_\vartheta$ for particles oriented at angle $\vartheta$).
Conversely, at high friction ($\mu_p \ge 1.0$), $\tilde{\omega}$ becomes large, exceeding even the viscous Jeffery prediction \cite{jeffery_motion_1922, bretherton_motion_1962}.
Although this resembles the injection of noise in viscous systems \cite{talbot_exploring_2024}, inter-particle friction increases the actual driving torque alongside the noise. Therefore, the system never simply reduces to the infinite-noise fluid limit, \emph{i.e.} uniform rotation with the background fluid.
This is the ``frictional gearing'' state: sliding is suppressed, and non-hydrodynamic contacts force particles to tumble (yellow curves).
As a simple explanation, consider the limit of infinitely slender particles.
If they are perfectly aligned with the flow, frictional particles cannot slide, resulting in a finite torque even for thin rods perfectly aligned with the flow, which explains why they can roll more than the Jeffery prediction.
This transition is well-described by a phenomenological sigmoid function (dashed lines, see Eq.~\ref{eq:two_state} in the SM~\ref{app:fit_rotation}~\cite{SupplementalMaterial}), that models the crossover from sliding to rolling contacts \cite{degiuli_phase_2016, nagy_flow_2020, bilotto_shear_2025a}.
using physically consistent fit parameters.

To quantify the balance between non-equilibrium driving and equilibrium-restoring effects, we derive a diagnostic dimensionless parameter $\Pi$.
This parameter, an effective Ericksen number \cite{gennes_physics_1995}, directly compares the characteristic rates of the competing fluxes in Eq.~ \ref{eq:full_smoluchowski3d}.
The non-equilibrium driving flux, which forces tumbling, is set by the phenomenologically-determined rotational driving, $\mathcal{J} \sim \langle \omega_z \rangle$.
The quasi-equilibrium restoring flux arises from the system's attempt to return to the nematic state $f_{\text{eq}}$, and scales as the rotational diffusion $D_r$ multiplied by the characteristic theoretical gradient of the steric potential, $|\nabla_{\mathbf{u}} \mathcal{U}_{ex}| \sim U_2 S_2^{\text{eq}}$, where $S_2^{\text{eq}}$ is the nematic order parameter from equilibrium theory and we neglected the smaller contribution of the $P_4$ term.
The ratio of these two competing rates, which is the ratio of the two governing dimensionless numbers, defines $\Pi$:
\begin{equation}
    \Pi = -\frac{\langle \omega_z \rangle}{D_r U_2 S_2^{eq}} \, .
\end{equation}
This parameter maps the system's state: the quasi-equilibrium ``screening'' regime corresponds to $\Pi \ll 1$, where restoring steric flux dominates, while the far-from-equilibrium ``gearing'' regime corresponds to $\Pi \gg 1$, where driving dominates.
Fig.~\ref{fig:Jeffery_comparison_pi}c shows that $\Pi < 1$ for $\mu_p \le 0.4$, and even in the infinite frictional limit, where virtually all the contacts are sticking, $\Pi$ stays finite and order one.
However, as shape anisotropy decreases, the system undergoes a transition yielding a close-to-isotropic state.
In this regime, the system is inherently non-equilibrium: there is no pre-existing order to screen the flow. 
The dynamics are no longer a balance of fluxes but are dictated solely by the driving strength, making the rotational P\'eclet number, $\text{Pe} = \langle \omega_z \rangle / D_r =  \Pi  \, U_2\, S_2$, the relevant control parameter.
This is evidenced by the consistently non-zero values of the nematic parameter observed even at very small elongations \cite{marschall_orientational_2019, marschall_sheardriven_2020}.
\begin{figure}
    \centering
    \includegraphics[width=\linewidth]{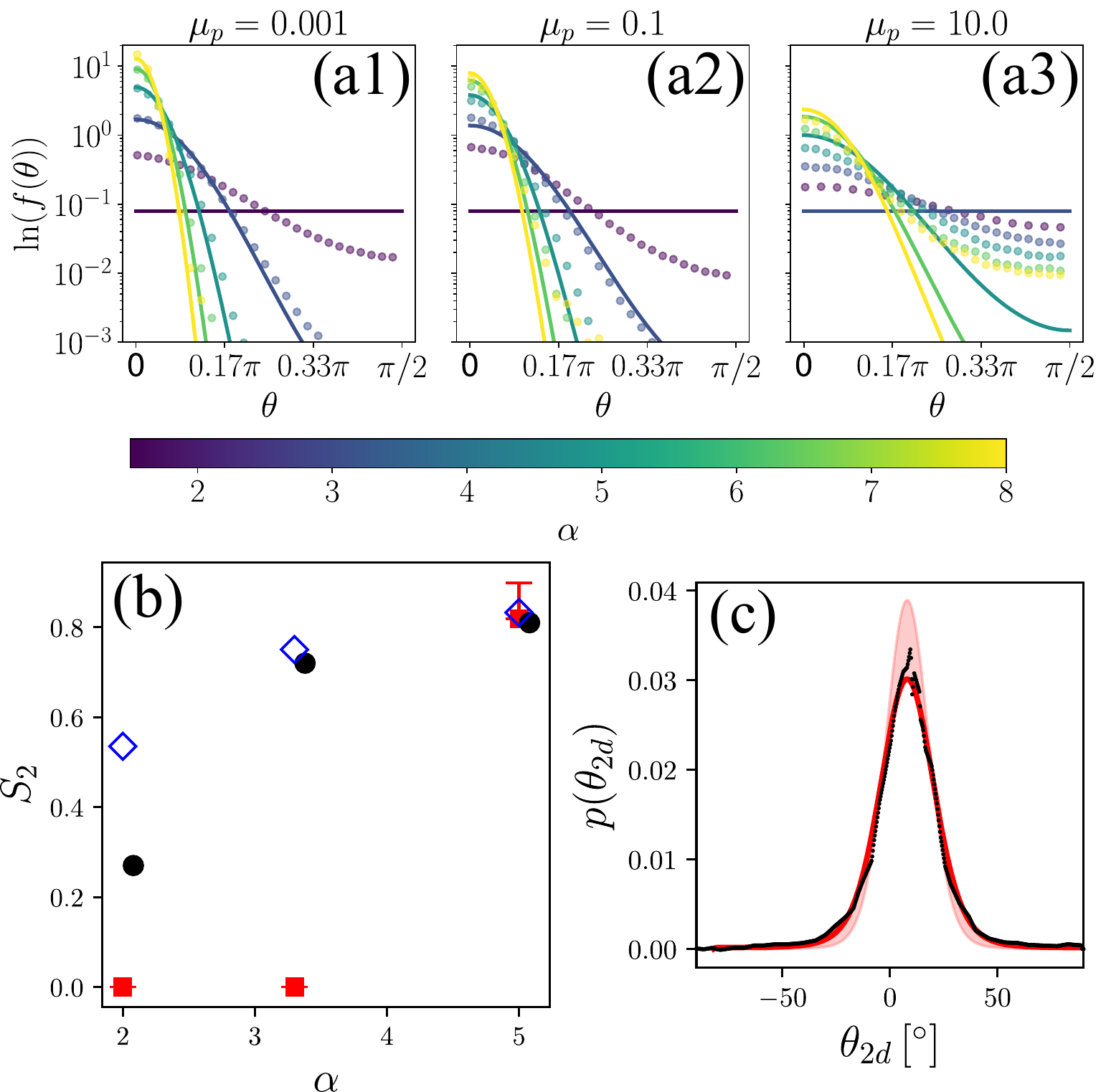}
    \caption{
    (a1-a3) Log-scale orientational distributions $f(\theta)$ (markers) compared to the equilibrium Maier-Saupe theory (Eq.~\ref{eq:eq_3D}, solid lines).
    The data reveals the breakdown of the equilibrium analogy as friction increases.
    (Left) In the low friction, screening regime ($\mu_p=0.001$), the theory accurately predicts the simulation data at high elongation, but fails at small $\alpha$.
    (Middle) In the intermediate friction case ($\mu_p=0.1$), the simulated distributions become noticeably broader than predicted, though the theory still qualitatively captures the overall trends. 
    (Right) In the high-friction, gearing regime ($\mu_p=10.0$), the theory fails, underestimating the broadening of the distribution even at high $\alpha$.
    Within each panel, colors represent aspect ratio $\alpha$, with higher $\alpha$ (yellow) leading to stronger ordering (sharper peaks).
    (b) Nematic order parameter $S_2$ as a function of aspect ratio $\alpha$, comparing experimental measurements \cite{wegner_alignment_2012} (black circles) with equilibrium theory predictions (red squares) and average for simple shear rod simulations with $\mu_p \in (0.1, 0.7)$, (blue hollow diamonds). 
    Theoretical points are shifted slightly along the horizontal axis for visual clarity. 
    At small elongations, the theory predicts an isotropic state ($S_2 \approx 0$) rather than the observed nematic distribution.
    Conversely, at high friction the order parameter is remarkably close to the one from equilibrium theory.
    At moderate and high elongation there is a great correspondence between split-bottom shear cell experiments and simple shear simulations.
    Red error bars represent the upper theoretical bounds assuming a $+10\%$ uncertainty (underestimation) in the experimentally measured volume fraction.
    (c) Comparison of the experimental in-plane orientation distribution $p(\theta_{2d})$ for $\alpha=5$ (black dotted line) \cite{wegner_alignment_2012} against a theoretical reconstruction (red solid line). 
    The reconstruction represents a 2D projection of the uniaxial 3D distribution estimated from the measured nematic order parameter, shifted to align with the experimentally measured average in-plane angle.
    The red shaded region reflects a plausible $+10\%$ uncertainty in the experimentally measured volume fraction due to binarization.
}
    \label{fig:f_theta}
\end{figure}

The breakdown of the equilibrium analogy is visualized in the orientational distributions (Fig.~\ref{fig:f_theta}a).
These distributions are measured around the director and limited to the polar angle, to focus on the breadth of the distribution (see SM~\ref{app:biaxiality}~\cite{SupplementalMaterial} for flow-induced anisotropy).
As expected, higher aspect ratios produce progressively sharper peaks.
However, increasing friction ($\Pi>1$) broadens the distributions, and the equilibrium Onsager--Parsons--Lee theory systematically fails, confirming the breakdown of the equilibrium analogy.
The continuous lines present the prediction from the Parsons--Lee equilibrium theory \cite{parsons_nematic_1979, lee_numerical_1987} (see SM~\ref{app:excluded_volume}~\cite{SupplementalMaterial}). 
At high aspect ratio and low friction, the data and the theory show excellent overlap.
However, these distributions reveal the two distinct failure modes of the equilibrium framework.
First, at low $\alpha$ ($\alpha < 2.5$), the equilibrium theory predicts an isotropic state ($S_2^{\text{eq}} \approx 0$) due to negligible steric interactions, whereas the simulation data (dots) reveals shear-induced nematic ordering driven by the non-equilibrium alignment term $\tilde{\mathbfcal{J}}$.
Second, at high friction ($\mu_p=10.0$) and high $\alpha$, the theory predicts a much sharper peak than observed.
Here, the non-equilibrium ``frictional gearing'' ($\Pi \gg 1$) actively disorders the system, broadening the distribution beyond the thermal prediction.
This broadening is sensitive to the inertial number $I$, unlike the robust quasi-equilibrium regime (see SM~\ref{app:rob_check}).

To confirm the {robustness} of these findings across different particle shapes and beyond ideal simulations, we carried out additional simulations of superellipsoidal rod-like particles (see SM~\ref{app:theory_experiments}~\cite{SupplementalMaterial}) and compared our framework directly against experimental measurements of granular rods from Ref.~\cite{wegner_alignment_2012}.
As shown in Fig.~\ref{fig:f_theta}b, at moderate and high elongations, there is a great correspondence between the split-bottom shear cell experiments and our simple shear rod simulations.
To directly compare the theory with experimental distributions, we reconstruct the theoretical equilibrium 3D distribution, using only the experimentally measured volume fraction $\phi$, and project it onto the 2D plane, which mathematically averages out the out-of-plane orientations to mimic the view of the experimental camera:
\begin{align}
    p(\theta_{2d}) = {1 \over \pi |\cos \theta_{2d}|} 
    \int_0^{\cos \theta_{2d}}  
    \frac{t f_{eq}(\arccos t )}{\sqrt{\cos^2\theta_{2d}-t^2}}dt \, ,
\end{align}
where $t= \cos{\theta}$.
As shown in Fig.~\ref{fig:f_theta} (bottom panels), the reconstructed $S_2$ mirrors the trends observed in simulations: the theory captures the strong alignment at high $\alpha$ but fails to predict the shear-induced ordering at low $\alpha$.
Crucially, for $\alpha=5.0$, the projected theoretical distribution shows excellent agreement with the experimental data, validating that the quasi-equilibrium description holds for real dissipative granular media in the steric screening regime.

\section{Conclusions and Outlook}

\begin{figure}
    \centering
    \includegraphics[width=0.8\linewidth]{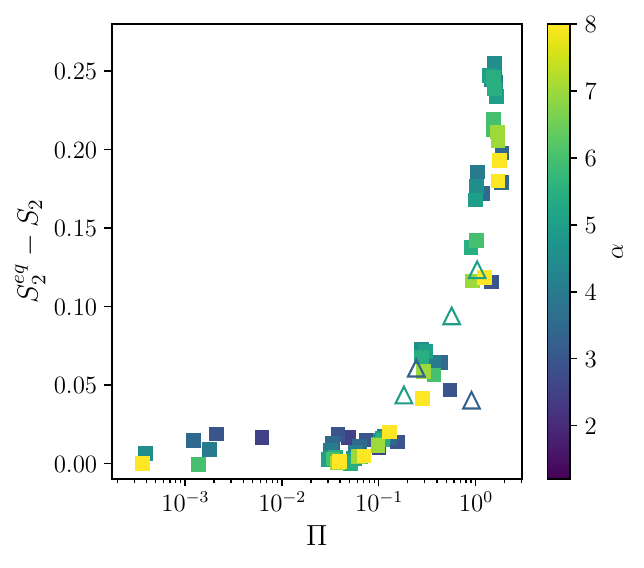}
    \caption{The positive deviation between the measured nematic order parameter and the equilibrium prediction, $S_2^{\text{eq}} - S_2$, plotted against the effective Ericksen number $\Pi$. 
    Data points include both spheroids (squares) and superellipsoidal rods (hollow triangles) simulations for different aspect ratios.
    The collapse confirms that $\Pi$ acts as a control parameter: for $\Pi \ll 1$ (steric screening), the system remains in a quasi-equilibrium state with minimal deviation, while for $\Pi \gtrsim 1$ (frictional gearing), the system is driven far from equilibrium, leading to a systematic divergence from the theoretical prediction.}
    \vspace{-.5cm}
    \label{fig:delta_s2}
\end{figure}

Finally, we utilize both the spheroid and rod simulation datasets to test our diagnostic metric $\Pi$.
In Fig. \ref{fig:delta_s2} we show the deviation of the equilibrium nematic order parameter from the one measured in simulations $S_2^{eq} - S_2$.
We only show positive deviations since negative ones are associated with low alpha values $\alpha \leq 2.5$, where the equilibrium alignment is mainly dictated by the shear.
The data shows a nice collapse as a function of $\Pi$, for both rods and ellipsoids.
At low $\Pi$ there is virtually no discrepancy, whereas as $\Pi$ increases there is a clear correlation with the distance from equilibrium.
This confirms that $\Pi$ is the correct metric to gauge distance from equilibrium for those grains elongated enough so that their distribution would be nematic in a thermal system.

In summary, we have shown that the orientational statistics of frictionless elongated granular particles under steady shear are quantitatively described by a mean-field Onsager-like theory, originally developed for thermal systems, when corrected via the Parsons--Lee scaling.
This confirms the existence of a ``quasi-equilibrium" nematic state in a driven dissipative system, proving that simple noisy Jeffery orbits are insufficient to model the dynamics once steric interactions become significant.
However, as inter-particle friction is introduced, the system is driven far from equilibrium, and the equilibrium analogy systematically breaks.
We show this breakdown manifests as a transition from a ``steric screening'' state to a ``frictional gearing'' state, where contacts actively promote particle rotation.
This transition is well-captured by a simple phenomenological model, whose parameters show clear physical trends with particle shape.
We quantify this with an effective Ericksen number, taken as the ratio of the driving over the restoring nematic torque, akin to active matter \cite{marchetti_hydrodynamics_2013} which captures the distance from equilibrium for sufficiently elongated grains. 
By leaving the specific microscopic driving flux as an agnostic input, our architecture remains decoupled from purely granular contact mechanics.
This allows the dimensionless breakdown diagnostic $\Pi$ to serve as a general blueprint for other dense far-from-equilibrium anisotropic systems. 
The framework established here provides a new, parameter-free tool to map out non-equilibrium ordering transitions not only in advanced kinetic granular theories \cite{vescovi_simple_2024, berzi_shaking_2025}, but also in dense driven suspensions \cite{trulsson_directional_2021, franceschini_transverse_2011} and active matter \cite{marchetti_hydrodynamics_2013}, defining the boundary where equilibrium analogies fail.

\section*{Author contributions}
Conceptualization: J.B., M.T., J.-F.M.,
Data curation: J.B., 
Formal analysis: J.B., M.T.
Funding acquisition: J.B., M.T., J.-F.M., 
Investigation: J.B.,
Methodology: J.B., M.T.
Software: J.B.,
Supervision: M.T., J.F.-M.
Visualization: J.B.,
Writing – original draft: J.B.,
Writing – review \& editing: M.T., J.F.-M. 

\section*{Conflicts of interest}

There are no conflicts to declare.

\section*{Data availability}

Data to support these findings are available on Zenodo (\url{https://doi.org/10.5281/zenodo.23163285}). Codes to run the simulations and the analysis are available on GitHub (\url{https://github.com/JBil8/Data_processing_orientations}).

\section*{Acknowledgements}
This project has received funding from the European Union’s Horizon 2020 research and innovation programme under the Marie Skłodowska-Curie grant agreement No 945363.
J.B. and J.-F.M. also acknowledge financial support from Bühler AG.
M.T. acknowledges funding from the Swedish Research Council under grant No. 2021-04997
We also acknowledge helpful discussions with Tamás Börzsönyi and Pierpaolo Bilotto.

\bibliography{references}
\appendix
\input{ESI}

\end{document}

%% file: ESI.tex
\clearpage
\onecolumngrid
\setlength{\columnwidth}{\textwidth}
\setlength{\linewidth}{\textwidth}
\setcounter{section}{0}
\setcounter{subsection}{0}
\setcounter{equation}{0}
\setcounter{figure}{0}
\setcounter{table}{0}
\renewcommand{\thesection}{\Alph{section}}
\renewcommand{\thesubsection}{\arabic{subsection}}
\renewcommand{\theequation}{S\arabic{equation}}
\renewcommand{\thefigure}{S\arabic{figure}}
\renewcommand{\thetable}{S\arabic{table}}
\section*{Supplementary material}
\section{Convergence to Non-Equilibrium Steady State}
\label{app:steady_state}

\begin{figure}[h]
    \centering
    \includegraphics[width=.5\linewidth]{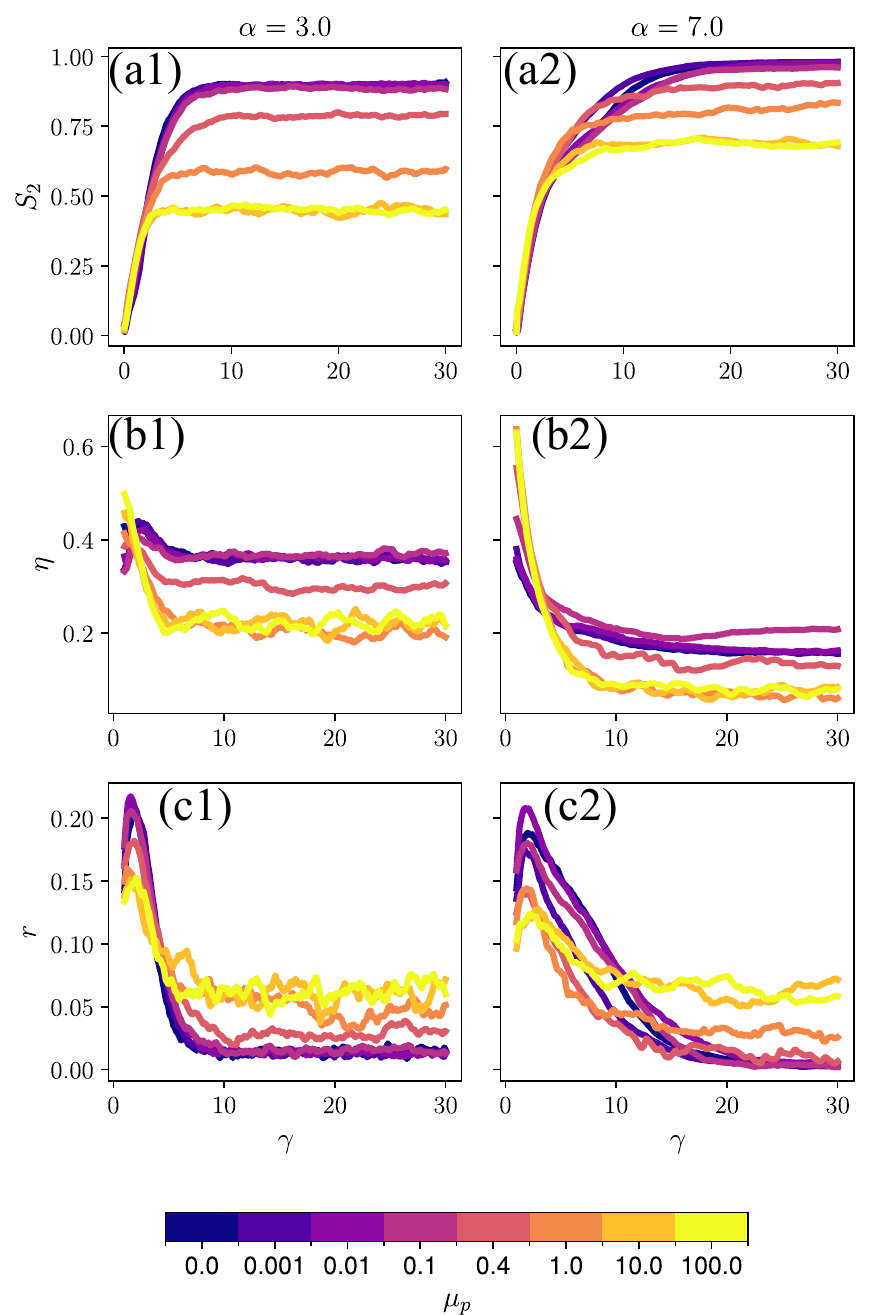}
    \caption{Transient evolution of (a) the nematic order parameter $S_2$, (b) the director angle $\eta$, and (c) the biaxiality $r$ as a function of accumulated strain $\gamma$ in simulations.
    Data is shown for aspect ratios $\alpha=3.0$ (left column) and $\alpha=7.0$ (right column).
    All systems converge to stable plateaus, confirming the absence of director tumbling.
    While increasing elongation slows down the dynamics, increasing friction (lighter colors) accelerates the convergence to the steady state.}
    \label{fig:S2_time}
\end{figure}

To validate our steady-state analysis, we demonstrate that all simulated systems converge to a Non-Equilibrium Steady State (NESS).
Fig.~\ref{fig:S2_time} tracks the evolution of the key orientational descriptors---$S_2$, the director angle $\eta$, and the biaxiality parameter $r$, see SM~\ref{app:biaxiality} for the definition---from an initially isotropic state.
Crucially, the director angle $\eta$ (Fig.~\ref{fig:S2_time}b) saturates to a constant value for all parameters. 
This stability explicitly rules out the persistent tumbling or wagging dynamics often observed in polymeric liquid crystals \cite{larson_arrested_1990}, confirming that dense granular shear flows operate strictly in the flow-aligning regime.
A conservative strain of $\gamma=20$ is sufficient to reach this steady state across all aspect ratios.

The transient dynamics reveal physical trends consistent with our theoretical framework.
More elongated particles ($\alpha=7.0$, right column) require larger strains to order than shorter ones ($\alpha=3.0$, left column), a consequence of their lower rotational mobility.
Notably, high-friction grains (yellow curves) reach the NESS significantly faster than their frictionless counterparts (dark curves).
This is consistent with the ``gearing" mechanism: the strong non-equilibrium driving flux ($\Pi \gg 1$) rapidly forces particles into their kinematic steady state, whereas low-friction systems rely on slower, diffusive relaxation to find their equilibrium configuration.
Furthermore, the steady-state plateaus for high-friction systems exhibit larger fluctuations.
This increased noise is attributed to both the stronger rotational torques generated by tangential contact forces and the looser packing (dilatancy), which weakens steric caging.

\section{Robustness Checks: Ensembles, Contact Mechanics, and Finite-Size Effects}
\label{app:rob_check}

To ensure the generality of our findings, we verify that the observed orientational statistics are robust across different thermodynamic ensembles, contact stiffnesses, system sizes, and inertial numbers.

\begin{figure}[h]
    \centering
    \includegraphics[width=0.7\linewidth]{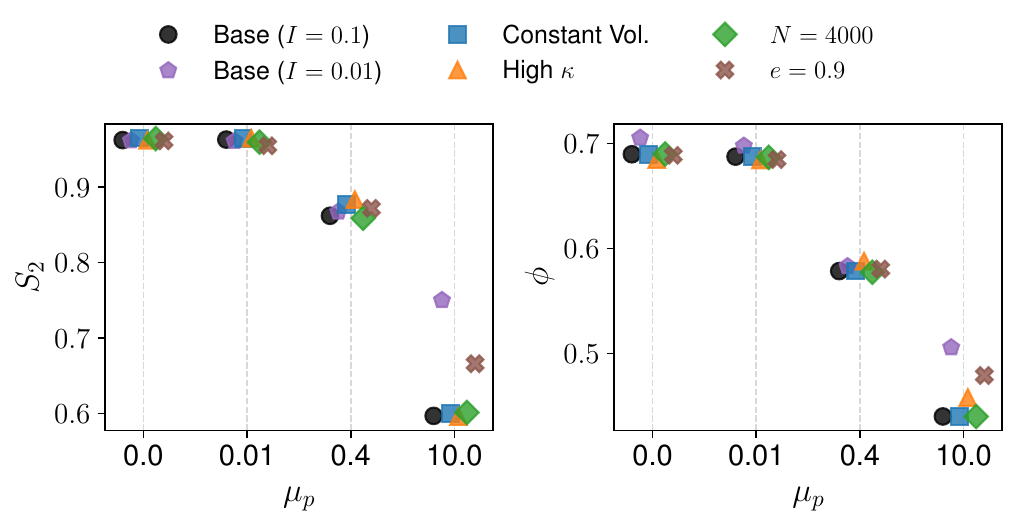}
    \caption{Robustness checks for the macroscopic steady-state properties.
    (Left) Nematic order parameter $S_2$ and (Right) packing fraction $\phi$ as a function of inter-particle friction $\mu_p$ for $\alpha=5.0$.
    The data compares the baseline simulations ($I=0.1$, black circles) against constant volume ensemble (blue squares), high contact stiffness (orange triangles), larger system size ($N=4000$ green diamonds), low inertial number ($I=0.01$, purple pentagons), and more elastic collisions ($e=0.9$, brown crosses). 
    For the same value of friction, data is shifted horizontally for ease of comparison.
    The collapse of the markers at low friction confirms the quasi-equilibrium state is completely robust to kinetic and ensemble parameters.}
    \label{fig:distribution_checks}
\end{figure}

Figure \ref{fig:distribution_checks} summarizes the macroscopic steady-state properties (nematic order $S_2$ and packing fraction $\phi$) across these variations for $\alpha=5.0$.

First, we compare simulations performed at constant normal stress ($\sigma_{yy}$, the standard protocol used in the main text) with those performed at constant volume ($V$).
For the constant volume runs, the packing fraction $\phi$ was fixed to the average steady-state value measured in the corresponding $\sigma_{yy}$ simulation.
The resulting $S_2$ values overlap.

Second, we verify that the observed order is not an artifact of soft-particle overlaps.
Increasing the contact stiffness to $\kappa = 4.6 \cdot 10^5$ (orange triangles) yields no significant deviation from the baseline, confirming we are in the hard-particle limit.

Third, we rule out finite-size effects by scaling the system up to $N=4000$ particles. 
When scaled isotropically (green diamonds), the system recovers the $N=2000$ results. 
We also tested an anisotropic, elongated box in the gradient direction.
However, at low friction, the elongated domain triggers weak shear localization,  where global spatial averages of $\phi$ and $S_2$ lose their thermodynamic meaning. 
Understanding the exact mechanism behind this geometry-induced localization remains an interesting topic for future studies.
We thus excluded this geometry from Fig.~\ref{fig:distribution_checks}. 
We conclude that, for uniform homogeneous shear, the $N=2000$ domain is able to capture bulk steady-state statistics.

\begin{figure}
    \centering
    \includegraphics[width=.5\linewidth]{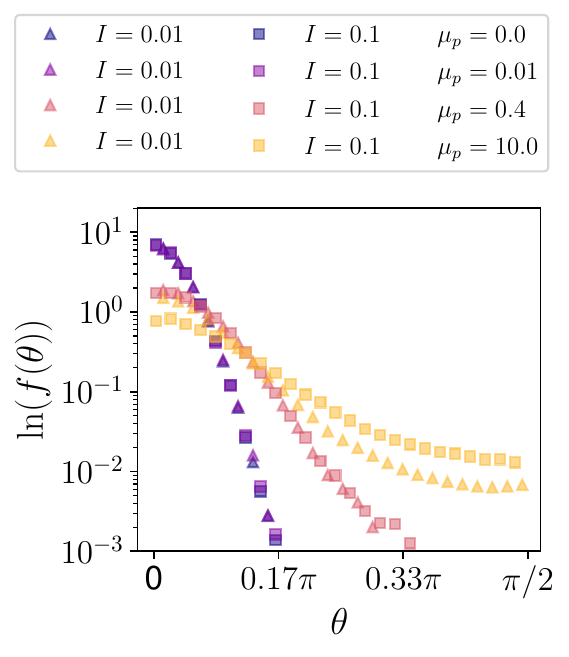}
    \caption{Comparison for inertial numbers $I=0.01$ (triangles) and $I=0.1$ (squares). 
    The quasi-equilibrium state ($\mu_p=0.0$) is robust to $I$, whereas the far-from-equilibrium gearing state ($\mu_p=10.0$) shows rate-dependence.
    Colors denote friction $\mu_p$.}
    \label{fig:I_comparison}
\end{figure}

Fourth, we test the sensitivity of the system to the inertial number $\mathcal{I}$.
Fig.~\ref{fig:I_comparison} compares distributions for $I=0.01$ (quasi-static limit) and $I=0.1$ (moderate flow).
In the low-friction ``screening'' regime (dark curves), the distributions are indistinguishable, supporting the idea that the system is in a robust quasi-equilibrium state governed essentially by geometry.
However, in the high-friction ``gearing'' regime (yellow curves), the distribution broadens noticeably as $I$ increases.
This sensitivity to shear rate is a hallmark of far-from-equilibrium dynamics, confirming that the gearing state lies outside the quasi-equilibrium regime.

Finally, we tested the influence of the normal restitution coefficient from the baseline ($e=0.1$) to a nearly elastic limit ($e=0.9$, brown crosses).
In the low-friction screening regime ($\mu_p \le 0.4$), the macroscopic state shows no significant deviation, further confirming that this quasi-equilibrium state is governed by steric exclusion rather than granular kinetics. 
Conversely, in the high-friction gearing regime ($\mu_p = 10.0$), the system becomes highly sensitive to $e$. This kinetic dependence, mirroring the sensitivity to the inertial number, physically validates that the gearing regime is firmly far-from-equilibrium.

\section{Validation of Rod Simulations}
\label{app:theory_experiments}

\begin{figure}
    \centering
    \includegraphics[width=\linewidth]{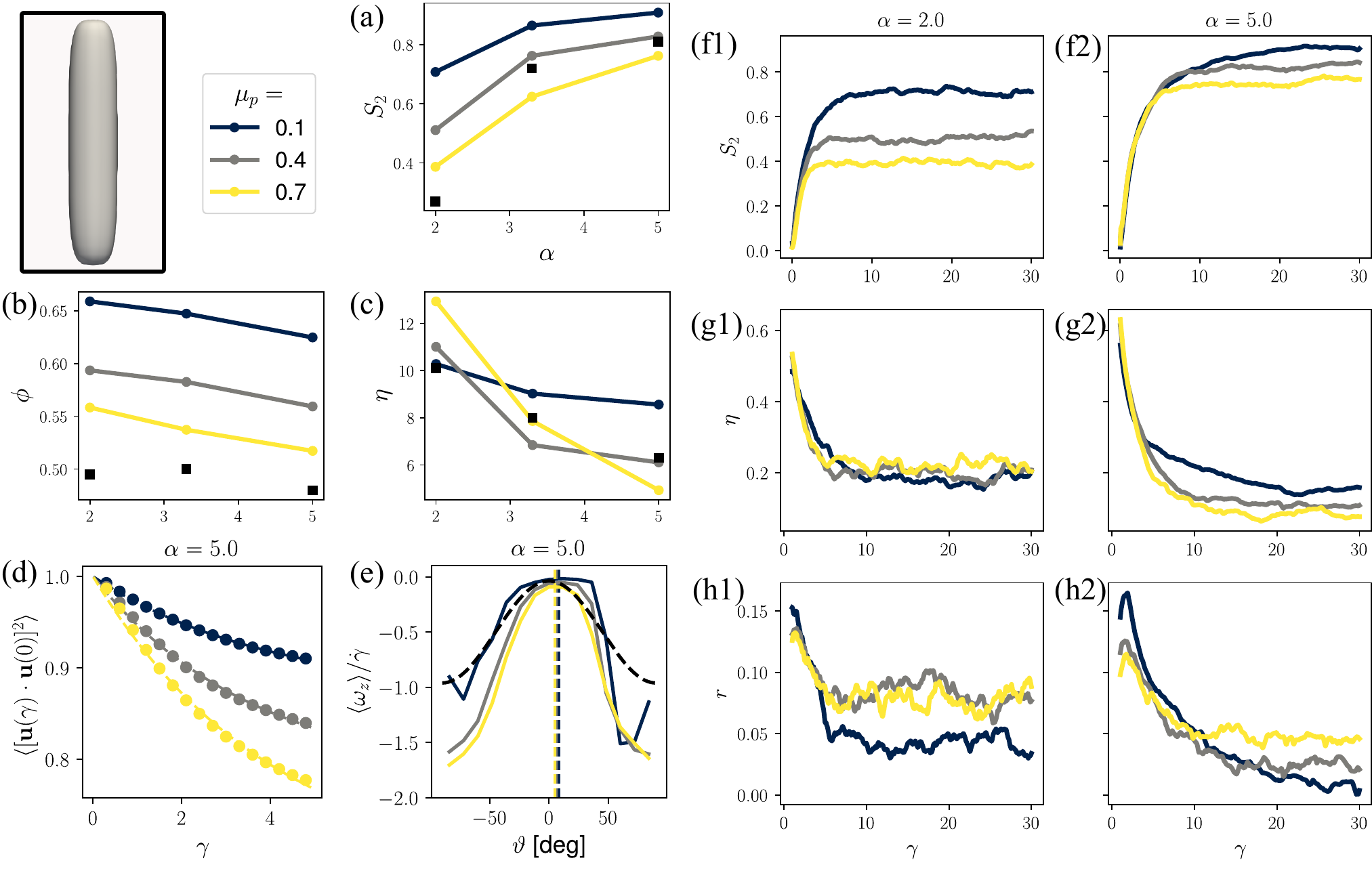}
    \caption{
    A rendering of the superellipsoidal particle shape used for $\alpha=5.0$ is shown in the top left box.
    (a) Nematic order parameter $S_2$, (b) packing fraction $\phi$ and (c) alignment angle $\eta$ vs. aspect ratio $\alpha$. 
    Black squares represent experimental data \cite{wegner_alignment_2012}; 
    lines are from rod simulations.
    (d) Decay of the orientational autocorrelation function.
    (e) Normalized average angular velocity $\langle\omega_z\rangle/\dot{\gamma}$ vs. angle $\vartheta$ for $\alpha=5.0$. 
    Vertical dashed lines mark the director angle.
    (f) Convergence of $S_2$, (g) $\eta$, and (h) biaxiality $r$ vs. strain $\gamma$ for $\alpha=2.0, 5.0$, showing steady state is reached by $\gamma \approx 20$.
    Colors correspond to $\mu_p=0.1$ (dark blue), $0.4$ (gray), and $0.7$ (yellow).
    }
    \label{fig:rods_supplementary}
\end{figure}

To validate our framework beyond spheroidal particles, we performed complementary Discrete Element Method (DEM) simulations using rod-like particles.
For an axisymmetric superellipsoidal particle, its level set in the principal axes is:
\begin{align}
   \left( {|\mathcal{X}|^{n_2} + |\mathcal{Y}|^{n_2} \over a^{n_2}} \right)^{n_1/n_2} + \left|{\mathcal{Z} \over c} \right|^{n_1} = 1    \, .
\end{align}
We set $n_2 = 2$ and $n_1 = 5$, yielding the superellipsoids shown in Fig.~\ref{fig:rods_supplementary} (top left).
With roughly 9\% less volume than a perfect cylinder of radius $a$ and length $2c$, this geometry serves as a proxy for the chamfered cylindrical rods studied experimentally by Wegner et al.~\cite{wegner_alignment_2012} at corresponding aspect ratios ($\alpha=2.0, 3.3, 5.0$).

Fig.~\ref{fig:rods_supplementary}a-c compares the steady-state nematic order parameter $S_2$, packing fraction $\phi$ and alignment angle $\eta$ obtained from our simulations with the experimental data \cite{wegner_alignment_2012, wegner_effects_2014}.
Using realistic inter-particle friction coefficients ($\mu_p \approx 0.4 - 0.7$), we observe quantitative agreement for $S_2$ and $\eta$ with the experiments for moderate to high aspect ratios.
At low aspect ratios, the sharp edges of the rods used in experiments compared to the smooth edges of our superellipsoidal particles, together with gravity secondary flows could be the source of discrepancy.
However, the packing fraction $\phi$ in our simulations is systematically higher than the experimentally reported values.
This discrepancy likely stems from the image binarization technique used to extract density from the experimental X-ray CT scans \cite{wegner_effects_2014}, which relies on setting a voxel threshold to distinguish solid volume from voids. 
The authors of that study noted the potential for systematic error in these measurements, which motivated our inclusion of an upper uncertainty bound ($+10\%$) for the theoretical predictions shown in Fig.~\ref{fig:f_theta}b-c of the main text.

Ultimately, the agreement of the orientational metrics justifies the use of rod simulations in the main text (Fig. \ref{fig:delta_s2}) to demonstrate the validity of the $\Pi$-scaling.

Furthermore, we confirm that the rod systems exhibit the same dynamical signatures as the spheroids.
The orientational autocorrelation functions (Fig.~\ref{fig:rods_supplementary}d) display the characteristic decay discussed in SM~\ref{app:rot_diffusion}, and the angular velocity profiles (Fig.~\ref{fig:rods_supplementary}e) confirm the existence of the ``steric screening'' mechanism, where the rotation rate vanishes at the director angle.
Fig.~\ref{fig:rods_supplementary}f-h shows the convergence to a stable, non-tumbling steady state.

\section{Rotational diffusion: noise and its breakdown}
\label{app:rot_diffusion}

\begin{figure}
    \centering
    \includegraphics[width=.6\linewidth]{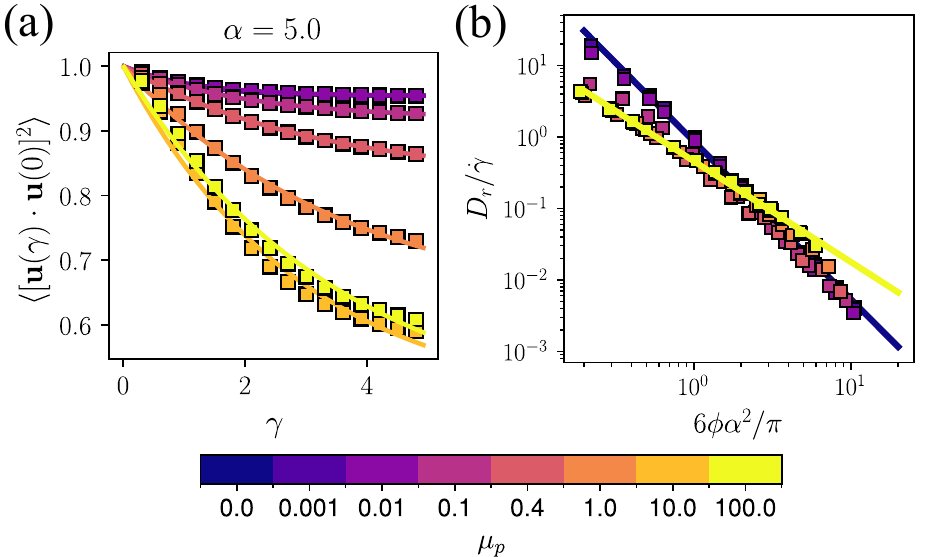}
     \caption{(a) Orientational autocorrelation function (OACF) vs. accumulated strain for $\alpha=5$. Dashed lines indicate fits to the nematic Brownian model (Eq.~\ref{eq:rot_diffusion_nematic}).
     (b) Dimensionless rotational diffusivity $\tilde{D}_r$ vs. normalized density. 
     Low-friction grains (dark markers) follow the inverse-square tube scaling (continuous dark line), confirming the quasi-equilibrium analogy. 
     High-friction grains' data (yellow markers) deviate, indicating the breakdown of the diffusive approximation in the gearing regime.
     }
    \label{fig:oacf_decay}
\end{figure}

Our ``quasi-equilibrium" hypothesis rests on the premise that shear-induced collisions in athermal systems act as an effective thermal noise.
We validate this by analyzing the orientational autocorrelation function (OACF), $\langle [\mathbf{u}(\gamma) \cdot \mathbf{u}(0)]^2 \rangle$.
In the low-friction ``screening" regime (Fig.~\ref{fig:oacf_decay}a, dark curves), the OACF collapses onto the single-exponential decay characteristic of nematic-caged Brownian dynamics \cite{chen_scaling_2020}:
\begin{align}
    \langle [\mathbf{u}(\gamma) \cdot \mathbf{u}(0)]^2 \rangle =
    A_\infty + (1-A_\infty) \, e^{-\gamma/\gamma_r} \,,
    \label{eq:rot_diffusion_nematic}
\end{align}
where $\gamma_r$ is the relaxation strain and $A_\infty$ is the long-time plateau arising from nematic caging, and function of $U_2$ and $S_2$.
This Brownian-like decay confirms that frequent, frictionless collisions effectively mimic a thermal bath.
The effective rotational diffusion coefficient is extracted directly from this decay:
\begin{align}
    \tilde{D}_r = \frac{D_r}{\dot \gamma} = \frac{1- A_\infty}{4 \gamma_r} \,.
\end{align}

To verify the physical origin of this diffusion, we compare our data to the extended tube model for thermal rods \cite{chen_scaling_2020}, which predicts the scaling:
\begin{align}
\frac{D_r}{D_r^0} \sim
\left(\frac{6 \phi \alpha^2}{\pi}\right)^{-2} g(\alpha, S_2)^{-2} ,
\label{eq:diffusion_volume}
\end{align}
where $D_r^0$ is the isolated-rod diffusivity (replaced here by $\dot{\gamma}$) and $g(\alpha, S_2) = (1+\pi/\alpha)+ (2\pi/\alpha - 1)S_2^2$ corrects for tube dilation.
Figure~\ref{fig:oacf_decay}b shows remarkable agreement between the theoretical scaling and our low-friction data (dark markers).
A power-law fit yields an exponent of $-2.2$, very close to the predicted $-2$, confirming that determining mobility via tube theory is valid for this granular system.

However, this analysis also captures the onset of the ``gearing" transition.
As friction increases (Fig.~\ref{fig:oacf_decay}a, yellow curves, $\mu_p \gg 1$), the OACF decays significantly faster and deviates from the simple Brownian form (Eq.~\ref{eq:rot_diffusion_nematic}).
Correspondingly, the measured diffusivity in Fig.~\ref{fig:oacf_decay}b (yellow markers) diverges from the tube-model scaling, following a weaker decay with an exponent of $\approx -1.41$.
This deviation is physical: $D_r$ is a quasi-equilibrium concept valid only when stochastic collisions dominate.
In the high-friction limit, the dynamics become deterministic and drive-dominated, rendering the diffusive description insufficient.

\section{Excluded volume computations}
\label{app:excluded_volume}

In the absence of a shear-driving flux ($\tilde{\mathbfcal{J}}=0$), the steady-state solution to Eq.~\ref{eq:full_smoluchowski3d} corresponds to the equilibrium state, $f_{\text{eq}}$, where the flux on the right hand side must also be zero:
\begin{equation}
\nabla_{\mathbf{u}} \cdot \left[ C(f) \left( \nabla_{\mathbf{u}} f_{\text{eq}} + f_{\text{eq}} \nabla_{\mathbf{u}} \mathcal{U}_{ex} \right) \right] = 0 \, .
\end{equation}
The solution to this yields a Boltzmann-like distribution governed by the mean-field steric potential:
\begin{equation}
f_{\text{eq}}(\mathbf{u}) = \frac{1}{Z} \exp(-\mathcal{U}_{ex}(\mathbf{u}))
\label{eq:S6}
\end{equation}
where $Z$ is a normalization constant. This justifies the Maier-Saupe form of Eq.~\ref{eq:eq_3D} in the main text.

The potential $\mathcal{U}_{\text{ex}}(\mathbf{u})$ accounts for the excluded volume interaction between a particle with orientation $\mathbf{u}$ and the mean-field of all other particles.
Following Onsager and Parsons \cite{onsager_effects_1949, parsons_nematic_1979}, it is expressed as:
\begin{align}
    \mathcal{U}_{\text{ex}}(\mathbf{u}) 
    = \mathcal{F}(\phi) \int d \mathbf{u}' f(\mathbf{u}') \, 
    \overline{V}_{\text{ex}}(\mathbf{u}, \mathbf{u}') \,,
    \label{eq:S7}
\end{align}
where $\phi$ is the packing fraction and $\overline{V}_{\text{ex}}=V_{\text{ex}}/V_p$ is the excluded volume normalized by the average particle volume $V_p$.
The prefactor $\mathcal{F}(\phi)$ is derived from the Carnahan--Starling correction for dense hard spheres \cite{parsons_nematic_1979, lee_numerical_1987}:
\begin{align}
     \mathcal{F}(\phi) = \frac{\phi (1 - 3\phi/4)}{(1 - \phi)^2} 
\end{align}

While more complex equations of state explicitly incorporating curvature terms exist, such as Vega's model for ellipsoids \cite{vega_virial_1997}, we found they provide negligible corrections to this baseline framework.
For axisymmetric particles, $\overline{V}_{\text{ex}}$ depends only on the relative angle $\theta$ and admits a Legendre expansion:
\begin{align}
    \overline{V}_{\text{ex}}(\theta) = \sum_{k=0, \text{even}}^{\infty} \overline{B}_k P_k(\cos\theta), 
\end{align}

Truncating at $k=4$, the effective interaction strengths in the Maier-Saupe potential (Eq.~2) are $U_{2,4} = - \mathcal{F}(\phi)\, \overline{B}_{2,4}$.

Exact coefficients for spheroids were derived in \cite{piastra_explicit_2015}, but we adopt the common Berne–Pechukas approximation \cite{berne_gaussian_1972}. Retaining the $P_4$ term yields:
\begin{align}
    \overline{V}_{\text{ex}}(\theta) &\approx 
    \overline{B}_2 P_2(\cos\theta) + \overline{B}_4 P_4(\cos\theta) \, , \\ 
    \overline{B}_2 &= -\frac{8}{3}\frac{\beta^2}{\sqrt{1-\beta^2}} 
    \left(1 + \frac{3\beta^2}{14}\right) \, , \\
    \overline{B}_4 &= -\frac{8}{35}\frac{\beta^4}{\sqrt{1-\beta^2}} 
    \, ,
\end{align}
where $\beta = (\alpha^2-1) / (\alpha^2+1)$ is the Bretherton parameter \cite{bretherton_motion_1962}.

For cylindrical rods (as in the experiments), the coefficients for $P_{2, 4}(\cos\theta)$ are \cite{onsager_effects_1949}:
\begin{align}
    \overline{B}_{2,r} = 
    -\frac{5\pi}{32}\left(\frac{8}{\pi}\alpha + \frac{2}{\alpha}\right) 
    + \frac{5}{4} + 0.385 \frac{8}{\pi}, \\
    \overline{B}_{4,r} = 
    -\frac{9\pi}{256}\left(\frac{8}{\pi}\alpha + \frac{2}{\alpha}\right) 
    - \frac{3}{8} - 0.065 \frac{8}{\pi} \, .
    \label{eq:second_coeff_rod}
\end{align}
We use this result to compute the theoretical interaction strengths $U_{2, 4}$ for the experimental data, using $\phi$ as measured from Wegner et al.~\cite{wegner_effects_2014}.
Given $\phi$, the equilibrium distribution $f_{\rm eq}(\mathbf{u})$ in Eq.~\ref{eq:S6} is obtained by solving the self-consistency condition (Eq.~\ref{eq:S7}) numerically:
writing $f$ in the ansatz $f \propto \exp[-\mathcal{F}(\phi)(\overline{B}_2 S_2 P_2 + \overline{B}_4 S_4 P_4)]$, the order parameters $S_2 = \langle P_2 \rangle$ and $S_4 = \langle P_4 \rangle$ are found by fixed-point iteration

\section{Effective Volume Fraction of Soft Hertzian Particles}
\label{app:softness_effects}

To quantify the influence of particle softness, we apply a perturbation approach \cite{barker_perturbation_1967} to the Parsons-Lee theory, analogous to the methods used for thermal soft spheres \cite{parsons_nematic_1979}.
Treating the applied normal stress as an effective energy scale $\sigma_{yy} d^3$ and comparing it to the Hertzian contact energy scale, $E d^3$, the dimensionless ratio dictating the perturbation is:
\begin{equation}
    \frac{\sigma_{yy}}{E} = 10^{-5}    
\end{equation}
The effective hard-sphere diameter $d_{\text{eff}}$ for a soft potential is defined by the integral:
\begin{equation}
    d_{\text{eff}} = \int_{0}^{d} \left( 1 - \exp\left[ - \frac{E }{\sigma_{yy}} \left(1 - \frac{r}{d} \right)^{(5/2)} \right] \right) dr    
\end{equation}
Evaluating the integral yields the first-order correction:
\begin{equation}
    \frac{d_{\text{eff}}}{d} \approx 1 - \Gamma\left(1 + \frac{2}{5}\right) \left(\frac{\sigma_{yy}}{E}\right)^{2/5} \, ,
\end{equation}
where $\Gamma$ is the gamma function.
Using $\Gamma(1.4) \approx 0.8873$ and our numerical ratio $P/E = 10^{-5}$:
\begin{equation}
    \frac{d_{\text{eff}}}{d} \approx 1 - (0.8873 \times 0.01) \approx 0.991    
\end{equation}
The effective volume fraction $\phi_{\text{eff}}$ scales with the cube of this diameter ratio:
\begin{equation}
    \frac{\phi_{\text{eff}}}{\phi} = \left(\frac{d_{\text{eff}}}{d}\right)^3 \approx (0.991)^3 \approx 0.974
\end{equation}
Assuming infinitely hard particles at this normal stress overpredicts the effective volume fraction by roughly $2.6\%$, confirming that the system is operating within the hard-particle limit.

\section{Phenomenological model parameters}
\label{app:fit_rotation}

The crossover from the quasi-equilibrium state to the drive-dominated state is captured by the phenomenological two-state model:
\begin{equation}
\tilde{\omega}= (1-S(\mu_p)) \tilde{\omega}_{ \mu_p \to 0} + S(\mu_p) \tilde{\omega}_{ \mu_p \to \infty}
\label{eq:two_state}
\end{equation}
where the sigmoid switch $S(\mu_p) = [1 + (\mu_p/\mu_t)^{-k}]^{-1}$ interpolates between the screened limit ($\tilde{\omega}_0 \equiv \tilde{\omega}_{ \mu_p \to 0}$) and the geared limit ($\tilde{\omega}_\infty \equiv \tilde{\omega}_{ \mu_p \to \infty}$).

\begin{figure}[]
    \centering
    \includegraphics[width=.4\linewidth]{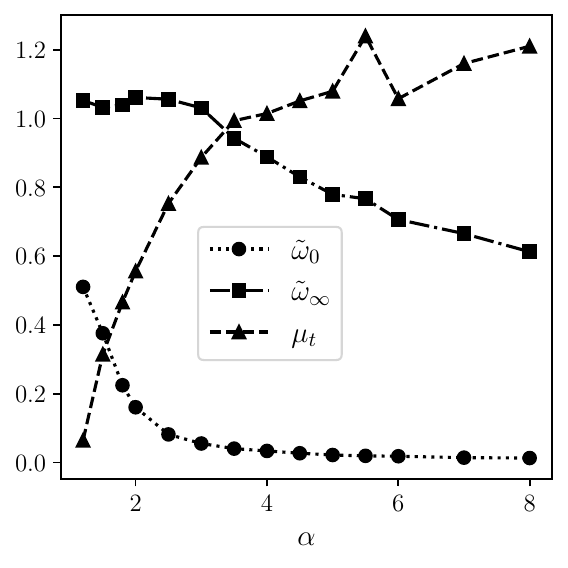} 
    \caption{Physical scaling of the phenomenological model parameters.
    The screened rotation $\tilde{\omega}_0$ vanishes for elongated particles, while the geared rotation $\tilde{\omega}_\infty$ remains finite.
    The transition friction $\mu_t$ scales with $\alpha$, indicating that stronger nematics are mechanically harder to break.
    The sharpness $k$ (not shown) is approximately constant $\approx 1.75$ across all shapes.}
    \label{fig:fit_params}
\end{figure}

Fig.~\ref{fig:fit_params} reveals the physical scaling of these parameters.
In the frictionless limit ($\tilde{\omega}_0$, circles), the rotation vanishes for $\alpha > 4.0$, confirming the ``steric screening'' mechanism where collective alignment effectively cancels the shear torque.
Conversely, the infinite-friction ``gearing'' limit ($\tilde{\omega}_\infty$, squares) remains finite but decreases with aspect ratio.
Crucially, this gearing rate is distinct from both hydrodynamic limits: it significantly exceeds the viscous Jeffery prediction, yet remains below the macroscopic fluid vorticity ($\tilde{\omega} = 1$).
This indicates that while friction enhances torque transmission, the granular assembly does not rotate as a rigid body.

The transition parameters offer further insight.
The critical friction $\mu_t$ (triangles) increases monotonically with $\alpha$: deeper steric barriers (stronger nematic order at high $\alpha$) require larger frictional forces to disrupt the equilibrium alignment.
Finally, the transition sharpness $k \approx 1.75$ is remarkably insensitive to particle shape. 
This constancy suggests a universal cooperative mechanism for the onset of tumbling, independent of the specific geometric details of the grains, but more investigation is needed to confirm this.

\section{Flow-Induced Anisotropy and Biaxiality}
\label{app:biaxiality}

\begin{figure}[]
    \centering
    \includegraphics[width=.5\linewidth]{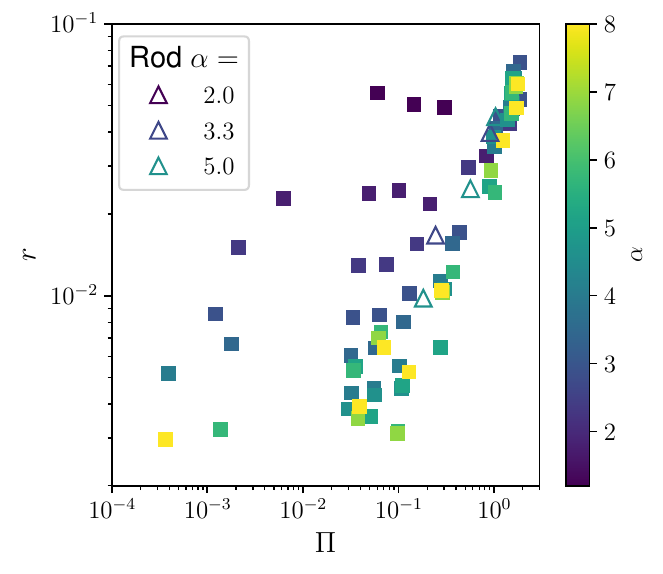}
    \caption{
    Biaxiality $r$ as a function of the effective Ericksen number $\Pi$.
    The deviation from uniaxiality ($r$) increases monotonically with $\Pi$.
    Data is shown for both spheroids (squares) and rods (triangles) with $\alpha > 2.5$.
    The collapse suggests that $\Pi$ controls the degree of non-equilibrium asymmetry.}
    \label{fig:exp_biax_angle}
\end{figure}

While the Maier-Saupe theory predicts a purely uniaxial alignment, the presence of shear flow inherently breaks this symmetry, inducing a biaxial state that serves as a key signature of non-equilibrium dynamics.
We quantify this deviation using the eigenvalues ($\lambda_1 \geq \lambda_2 \geq \lambda_3$) of the nematic tensor $\mathbf{Q}=\int_{\mathbb{S}^2} (\mathbf{u} \otimes \mathbf{u} - \mathbf{I}/3) f(\mathbf{u}) d\mathbf{u}$.
Its corresponding eigenvectors $\mathbf{e}_1$ and $\mathbf{e}_2$ lie in the flow-gradient plane, where the director angle $\eta$ is measured between $\mathbf{e}_1$ and $\mathbf{e}_x$, while $\mathbf{e}_3 = \mathbf{e}_z $.
The biaxiality parameter is defined as $r = \lambda_2 - \lambda_3$, where $r=0$ corresponds to a perfectly uniaxial distribution.
Fig.~\ref{fig:exp_biax_angle} demonstrates a strong link between this non-equilibrium signature and our diagnostic parameter $\Pi$.
For elongated particles ($\alpha > 4.0$), the biaxiality $r$ collapses onto an increasing function of $\Pi$, with minimal scattering across aspect ratios.
Crucially, this trend holds for both spheroids and rods.
This confirms that $\Pi$ successfully quantifies the system's distance from equilibrium: as $\Pi$ grows, the non-equilibrium driving flux increasingly dominates the restoring steric potential, pushing the system further from the uniaxial reference state.

\begin{figure}[h]
    \centering
    \includegraphics[width=.5\linewidth]{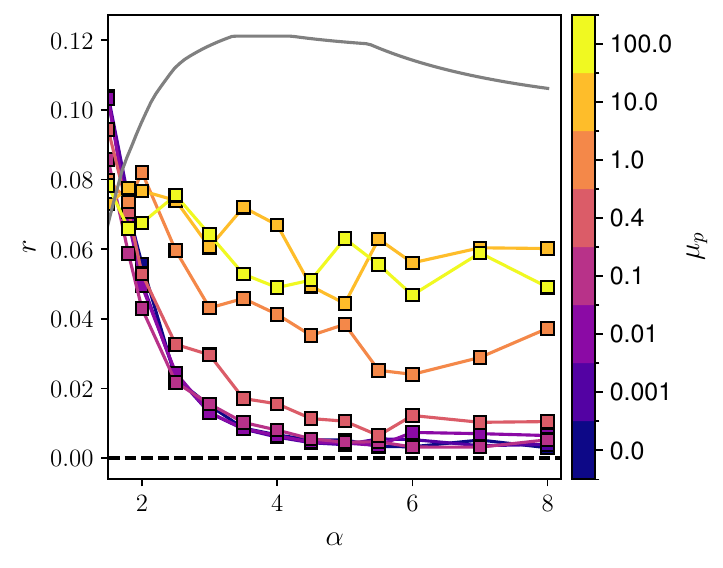}
    \caption{Biaxiality parameter $r$ as a function of aspect ratio $\alpha$.
        The dashed line marks the ideal uniaxial limit ($r=0$), while the gray line represents the theoretical prediction for Jeffery orbits \cite{talbot_exploring_2024}.
        Notably, the dense granular flow exhibits significantly lower biaxiality than the Jeffery prediction for $\alpha > 2$, implying that steric interactions actively suppress flow-induced asymmetry regardless of the regime.}
    \label{fig:biax_alpha}
\end{figure}

The dependence on particle shape and friction is detailed in Fig.~\ref{fig:biax_alpha}.
In the low-friction ``screening" regime (dark curves), $r$ remains minimal and close to zero, significantly below the theoretical prediction for Jeffery orbits without noise \cite{talbot_exploring_2024}.
This suppression of biaxiality is consistent with the ``quasi-equilibrium" hypothesis, where the system is effectively caged in a uniaxial potential.
However, as $\mu_p$ increases (yellow curves), $r$ grows significantly.
This confirms that the friction-induced "gearing" mechanism actively drives the system away from the uniaxial limit, generating the finite biaxiality.
Nevertheless, the biaxiality never reaches the limit of pure Jeffery orbits, not even in the gearing regime.

\section{Quantifying the Deviation: Jensen-Shannon Divergence}
\label{sec:js_div}

To provide a global map of the theory's accuracy across the parameter space $(\alpha, \mu_p)$, we compute the Jensen-Shannon divergence (JSD) between the simulated and theoretical orientational distributions.
Unlike the Kullback-Leibler divergence, the JSD is symmetric and bounded, making it an ideal metric for comparing probability distributions.
It is defined as:
\begin{align}
JSD(\tilde{f}_{\text{theory}}||\tilde{f}_{\text{simul}}) = \frac{1}{2} \biggl[D_{\text{KL}}
    \left(\tilde{f}_{\text{theory}}|| m\right)  + \nonumber \\ 
    D_{\text{KL}}\left(\tilde{f}_{\text{simul}}|| m\right)   
    \biggr] \, ,
\end{align}
where $m = (\tilde{f}_{\text{theory}} + \tilde{f}_{\text{simul}})/2$ is the mixture distribution and $D_{\text{KL}}(p||q) = \int p(\theta) \log_2 (p(\theta)/q(\theta))d\theta$.
The distributions are normalized over the polar angle interval $[0, \pi/2]$, such that $\tilde{f}(\theta) \propto f(\theta)\sin(\theta)$ accounts for the solid angle phase space.
Using base-2 logarithms ensures that $0 \leq \sqrt{JSD} \leq 1$.

\begin{figure}[]
    \centering
    \includegraphics[width=.5\linewidth]{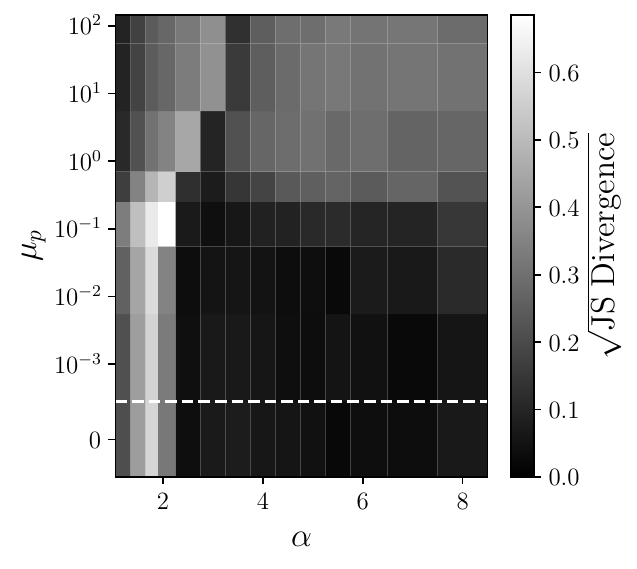}
    \caption{Map of theoretical validity. 
    The square root of the Jensen-Shannon Divergence ($\sqrt{JSD}$) is plotted as a function of friction $\mu_p$ and aspect ratio $\alpha$.
    Dark regions indicate excellent agreement between the Parsons--Lee equilibrium theory and DEM simulations.
    Bright regions indicate significant deviation, identifying the two distinct breakdown mechanisms: shear-dominance at low $\alpha$ and frictional gearing at high $\mu_p$.}
    \label{fig:JSD}
\end{figure}

The resulting distance metric is plotted in Fig.~\ref{fig:JSD}. 
The heatmap clearly identifies the quasi-equilibrium region: the error is negligible (dark regions) for $\alpha > 2.5$ and $\mu_p \leq 0.1$.
Outside this region, either due to the gearing transition at high friction or the dominance of shear alignment at low aspect ratio, the divergence grows rapidly (bright regions), quantitatively marking the breakdown of the equilibrium analogy.

%